\documentclass[preprint,aps,12pt,showpacs,nofootinbib,tightenlines]{revtex4}
\usepackage{amsmath}
\usepackage{amssymb}
\usepackage{epsfig}
\usepackage{graphicx}
\begin{document}

\newcommand{\beq}{\begin{eqnarray}}
\newcommand{\eeq}{\end{eqnarray}}
\newcommand{\non}{\nonumber\\ }

\newcommand{\acp}{{\cal A}_{CP}}
\newcommand{\etap}{\eta^{(\prime)} }
\newcommand{\etapr}{\eta^\prime }
\newcommand{\jpsi}{ J/\Psi }
\newcommand{\kst} {K^{*0}}
\newcommand{\kstb}{\overline{K}^{*0}}

\newcommand{\pb}{\phi_{B}}
\newcommand{\pjL}{\phi_{2}^L}
\newcommand{\pjt}{\phi_{2}^t}
\newcommand{\pjT}{\phi_{2}^T}
\newcommand{\pjV}{\phi_{2}^V}
\newcommand{\p}{\phi_{3}}
\newcommand{\pms}{\phi_{3}^s}
\newcommand{\pmt}{\phi_{3}^t}
\newcommand{\pmv}{\phi_{3}^v}
\newcommand{\pmT}{\phi_{3}^T}
\newcommand{\pma}{\phi_{3}^a}
\newcommand{\fb}{f_{B_s} }
\newcommand{\fpi}{f_{\pi} }
\newcommand{\fj}{f_{J/\Psi} }
\newcommand{\fetap}{f_{\eta'} }
\newcommand{\rpi}{r_{\pi} }
\newcommand{\rp}{r_{3} }
\newcommand{\rj}{r_{2} }
\newcommand{\mb}{m_{B_s} }
\newcommand{\mop}{m_{0\pi} }
\newcommand{\moe}{m_{0\eta} }
\newcommand{\moep}{m_{0\eta'} }
\newcommand{\ov}{\overline}

\newcommand{\psl}{ P \hspace{-2.8truemm}/ }
\newcommand{\nsl}{ n \hspace{-2.2truemm}/ }
\newcommand{\vsl}{ v \hspace{-2.2truemm}/ }
\newcommand{\epsl}{\epsilon \hspace{-1.8truemm}/\,  }

\def \epjc{ Eur. Phys. J. C }
\def \jpg{  J. Phys. G }
\def \npb{  Nucl. Phys. B }
\def \plb{  Phys. Lett. B }
\def \pr{  Phys. Rep. }
\def \prd{  Phys. Rev. D }
\def \prl{  Phys. Rev. Lett.  }
\def \zpc{  Z. Phys. C  }
\def \jhep{ J. High Energy Phys.  }
\def \ijmpa { Int. J. Mod. Phys. A }

\title{ Branching ratios of $B_c \to A P $ decays in the perturbative QCD approach}
\author{Xin Liu\footnote{liuxin.physics@gmail.com} and Zhen-Jun Xiao\footnote{xiaozhenjun@njnu.edu.cn}}
\affiliation{ Department of
Physics and Institute of Theoretical Physics, Nanjing Normal
University, Nanjing, Jiangsu 210046, Peoples' Republic of China }
\date{\today}
\begin{abstract}
In this paper we calculate the branching ratios (BRs) of
the 32 charmless hadronic $B_c \to A P$ decays ($A=a_1(1260),b_1(1235),K_1(1270),K_1(1400),
f_1(1285),f_1(1420),h_1(1170),h_1(1380)$) by employing the perturbative
QCD(pQCD) factorization approach.
These considered decay channels can only occur via annihilation type diagrams
in the standard model.
From the numerical calculations and phenomenological analysis, we found the
following results:
(a) the pQCD predictions for the BRs of the considered $B_c$
decays are in the range of $10^{-6}$ to $10^{-8}$, while the CP-violating
asymmetries are absent because only one type tree operator is involved here;
(b) the BRs of  $\Delta
S= 0$ processes are generally much larger than those of $\Delta S =1$
ones due to the large CKM
factor of $|V_{ud}/V_{us}|^2\sim 19$;
(c) since the behavior for $^1P_1$ meson is much different from
that of $^3P_1$ meson, the BRs of $B_c \to A(^1P_1) P$ decays
are generally larger than that of $B_c \to A(^3P_1) P$ decays;
(d) the pQCD predictions for the BRs of $B_c \to (K_1(1270), K_1(1400)) \etap $ and $(K_1(1270), K_1(1400)) K$
decays are rather sensitive to the value of the mixing angle $\theta_K$.
\end{abstract}


\maketitle

\section{Introduction}

Unlike the ordinary light $B_q$ ($q=u,d,s$) mesons, the $B_c$ meson is the only heavy
meson consisting of two heavy quarks $b$ and $c$  and
plays a special role in the precision test of the standard model(SM)~\cite{nb04:bcre}.
Moreover, a large number of $B_c$
meson events will be collected with the running of Large Hadron
Collider(LHC) experiments and this will provide great opportunities
for both theorists and experimentalists to study the perturbative and nonperturbative
QCD dynamics, final state interactions, etc.

In two recent works~\cite{ekou09:ncbc,Liu09:bcpv}, the pure annihilation
$B_c \to PP, PV/VP, VV$ decays (here $P$ and $V$ stand for the light pseudoscalar and vector
mesons) have been studied by employing the SU(3) flavor symmetry and
the pQCD  factorization approach ~\cite{Keum01:kpi,Lu01:pipi,Li03:ppnp}, respectively.

In the present work, we will study the two body charmless hadronic $B_c \to A P$ decays
(here $A$ denotes the light axial-vector mesons),
which can only occur via annihilation type diagrams in the SM.
First of all, the size of annihilation contributions is an important
issue in the $B$ meson physics, and has been studied extensively, for example, in
Refs.~\cite{Lu03:dsk,Keum01:kpi,Lu01:pipi,Hong06:direct,Li05:kphi,Gritsan07:kphi}.
Secondly, the internal structure of the axial-vector
mesons has been one of the hot topics in recent years \cite{Lipkin77,Yang05,Yang07:twist}.
Although many efforts on both
theoretical and experimental sides have been made
\cite{Cheng07:ap,Yang07:ba1,Wang08:a1,Cheng08:aa,Li09:afm,Amsler08:pdg,Barberio08:hfag}
to explore it through the  studies for the relevant decay rates, the CP-violating asymmetries,
polarization fractions and the form factors, etc., we currently still know little about the
nature of the axial-vector mesons.

In the quark model, there are two different types of light axial
vector mesons: $^3P_1$ and $^1P_1$, which carry the quantum
numbers $J^{\rm PC}=1^{++}$ and $1^{+-}$, respectively. The
$1^{++}$ nonet consists of $a_1(1260)$, $f_1(1285)$, $f_1(1420)$
and $K_{1A}$, while the $1^{+-}$ nonet has $b_1(1235)$,
$h_1(1170)$, $h_1(1380)$ and $K_{1B}$\footnote{For the sake of simplicity,
we will adopt the forms $a_1$ and $b_1$ to denote the non-strange axial-vector mesons
$a_1(1260)$ and $b_1(1235)$, respectively, in the following section. We will also use
$K_1$ to denote $K_1(1270)$ and $K_1(1400)$ for convenience unless otherwise stated.}.
In the SU(3) limit, these mesons can not mix with each other.
Because the $s$ quark is heavier than $u,d$ quarks, the meson $K_1(1270)$ and $K_1(1400)$
are not purely $1^3P_1$ or $1^1P_1$ state, but a mixture of $K_{1A}$ ($^3P_1$ state)
and $K_{1B}$ ( $^1P_1$ state).
Analogous to $\eta-\eta^\prime$ system, the flavor-singlet and flavor-octet
axial-vector mesons can also mix with each other. It is worth mentioning that
the mixing angles can be determined by the relevant data, but unfortunately, there is
no enough data now for these mesons which leaves the mixing angles basically free parameters.

In this paper, we will calculate the branching ratios of
the 32 non-leptonic charmless $B_c \to AP$ decays by employing
the low energy effective Hamiltonian~\cite{Buras96:weak} and the
pQCD factorization approach based on the framework of $k_T$
factorization theorem. By keeping the transverse momentum $k_T$ of
the quarks, the pQCD approach is free of endpoint singularity and
the Sudakov formalism makes it more self-consistent. In the pQCD approach
one can do the quantitative calculations of the annihilation type diagrams directly,
which can be seen, for instance, in
Refs.~\cite{Keum01:kpi,Lu01:pipi,Lu03:dsk,Li05:kphi,Liu09:bcpv}.

The pure annihilation $B_c\to PP, PV/VP, VV$ decays and $B_c\to A P$ decays considered
in Refs.~\cite{ekou09:ncbc,Liu09:bcpv} and in this paper generally have very small
branching ratios: at the order of $10^{-6}$ to $ 10^{-9}$.
According to the discussions as given in Ref.~\cite{ekou09:ncbc},
the charmless hadronic $B_c$ decays with decay rates at the level of
$10^{-6}$ could be measured at LHC experiments with the accuracy required for the
phenomenological analysis, while it may be difficult to
measure those $B_c$ decays if their branching  ratios are much less than $10^{-6}$.

The paper is organized as follows. In Sec.~\ref{sec:1}, we present
the formalism of the considered $B_c$ meson decays.
Then we perform the analytic calculations for considered
decay channels by using the pQCD approach in Sec.~\ref{sec:2}.
The numerical results and phenomenological analysis are given in
Sec.~\ref{sec:3}. Finally, Sec.~\ref{sec:sum} contains a short summary
and some discussions.

\section{Formalism}\label{sec:1}

In the pQCD approach, the decay amplitude of the two body decay $B_c \to M_1 M_2$
($M_1, M_2$ stand for the two final state mesons) can be written conceptually
as the convolution,
\beq
{\cal A}(B_c \to M_1 M_2)\sim \int\!\! d^4k_1 d^4k_2 d^4k_3\ \mathrm{Tr}
\left [ C(t) \Phi_{B_c}(k_1) \Phi_{M_1}(k_2) \Phi_{M_2}(k_3) H(k_1,k_2,k_3, t)
\right ],
\label{eq:con1}
\eeq
where $k_i$'s are momenta of light
quarks included in each mesons, and $\mathrm{Tr}$ denotes the
trace over Dirac and color indices. $C(t)$ is the Wilson coefficient
which results from the radiative corrections at short distance. In
the above convolution, $C(t)$ includes the harder dynamics at
larger scale than $m_{B_c}$ scale and describes the evolution of local
$4$-Fermi operators from $m_W$ (the $W$ boson mass) down to
$t\sim\mathcal{O}(\sqrt{\bar{\Lambda} m_{B_c}})$ scale, where
$\bar{\Lambda}\equiv m_{B_c} -m_b$. The function $H(k_1,k_2,k_3,t)$ describes the
four quark operator and the spectator quark connected by
 a hard gluon whose $q^2$ is in the order
of $\bar{\Lambda} m_{B_c}$, and includes the
$\mathcal{O}(\sqrt{\bar{\Lambda} m_{B_c}})$ hard dynamics. Therefore,
this hard part $H$ can be perturbatively calculated. The function $\Phi_M$ is
the wave function which describes hadronization of the quark and
anti-quark to the meson $M$.
In the present work, since the $B_c$ meson is composed of two heavy quarks
$b$ and $c$, we will take the nonrelativistic approximation form
$\delta(x- m_c/m_{B_c})$~\cite{DABc} for the distribution amplitude $\phi_{B_c}(x)$.
For light meson $A$ and $P$, we adopt the light-cone distribution amplitudes
directly, which will be displayed in Appendix~\ref{sec:app1}.
 While the function $H$ depends on the
processes considered, the wave function $\Phi_M$ is  independent of the specific
processes. Using the wave functions determined from other well measured
processes, one can make quantitative predictions here.

Since the b quark is rather heavy, we work in the frame with the
$B_c$ meson at rest, i.e., with the $B_c$ meson momentum
$P_1= (m_{B_c}/\sqrt{2})(1,1,{\bf 0}_T)$ in the light-cone
coordinates. For the charmless hadronic $B_c \to AP$ decays,
we assume that the $A$ ($P$) meson moves in the plus(minus) $z$
direction carrying the momentum $P_2$ ($P_3$), and with
the polarization vector $\epsilon_2$ for the $A$ meson.
Then the two final state meson momenta can be
written as
\beq
     P_2 =\frac{m_{B_c}}{\sqrt{2}} (1,r_A^2,{\bf 0}_T), \quad
     P_3 =\frac{m_{B_c}}{\sqrt{2}} (0,1-r_A^2,{\bf 0}_T),
\eeq
respectively, where $r_A=m_{A}/m_{B_c}$ and the mass of light
pseudoscalar mesons ($K, \pi$ and $\etap$) has been neglected.
For the axial-vector meson $A$, its longitudinal polarization vector, $\epsilon_2^L$,
can be defined as
\beq
\epsilon_2^L =\frac{m_{B_c}}{\sqrt{2}m_{A}} (1,
-r_A^2,{\bf 0}_T);
\eeq
Putting the (light-)  quark momenta in $B_c$, $A$ and $P$ mesons as $k_1$,
$k_2$, and $k_3$, respectively, we can choose
\beq
k_1 = (x_1P_1^+,0,{\bf k}_{1T}), \quad k_2 = (x_2 P_2^+,0,{\bf k}_{2T}), \quad
k_3 = (0, x_3 P_3^-,{\bf k}_{3T}).
\eeq
Then, for $B_c \to AP$
decays, the
integration over $k_1^-$, $k_2^-$, and $k_3^+$ 
will lead to the decay amplitudes in the pQCD approach,
\beq
{\cal A}(B_c \to AP) &\sim &\int\!\! d x_1 d x_2 d x_3 b_1
d b_1 b_2 d b_2 b_3 d b_3 \non && \cdot \mathrm{Tr} \left [ C(t)
\Phi_{B_c}(x_1,b_1) \Phi_{A}(x_2,b_2) \Phi_{P}(x_3, b_3) H(x_i,
b_i, t) S_t(x_i)\, e^{-S(t)} \right ]\;
\label{eq:a2}
\eeq
where $b_i$ is the conjugate space coordinate of $k_{iT}$, and $t$ is the
largest energy scale in function $H(x_i,b_i,t)$. The large
logarithms $\ln (m_W/t)$ are included in the Wilson coefficients
$C(t)$. The large double logarithms ($\ln^2 x_i$) are summed by the
threshold resummation ~\cite{Li02:resum}, and they lead to
$S_t(x_i)$ which smears the end-point singularities on $x_i$. The
last term, $e^{-S(t)}$, is the Sudakov form factor which suppresses
the soft dynamics effectively ~\cite{Li98:soft}. Thus it makes the
perturbative calculation of the hard part $H$ applicable at
intermediate scale, i.e., $m_{B_c}$ scale. We will calculate
analytically the function $H(x_i,b_i,t)$ for the considered decays
at leading order(LO) in $\alpha_s$ expansion and give the convoluted
amplitudes in next section.

For these considered decays, the related weak effective
Hamiltonian $H_{eff}$~\cite{Buras96:weak} is given by
\beq
\label{eq:heff}
H_{eff} = \frac{G_{F}} {\sqrt{2}} \, \left[
V_{cb}^* V_{uD} \left (C_1(\mu) O_1(\mu) + C_2(\mu) O_2(\mu)
\right) \right] \;\label{heff} ,
\eeq
with the current-current operators $O_{1,2}$,
\beq
O_1 &=& \bar u_\beta \gamma^\mu (1-\gamma_5)
D_\alpha \bar c_\beta \gamma^\mu (1- \gamma_5) b_\alpha \; , \non
O_2 &=& \bar u_\beta \gamma^\mu (1- \gamma_5) D_\beta \bar
c_\alpha \gamma^\mu (1- \gamma_5) b_\alpha \; ,
\eeq
where $V_{cb}, V_{uD}$ are the Cabibbo-Kobayashi-Maskawa (CKM) matrix elements, "D" denotes the
light down quark $d$ or $s$, and $C_i(\mu)$ are Wilson coefficients
at the renormalization scale $\mu$. For the Wilson coefficients
$C_{1,2}(\mu)$, we will also use the leading order (LO)
expressions, although the next-to-leading order calculations
already exist in the literature~\cite{Buras96:weak}. This is the
consistent way to cancel the explicit $\mu$ dependence in the
theoretical formulae. For the renormalization group evolution of
the Wilson coefficients from higher scale to lower scale, we use
the formulae as given in Ref.~\cite{Lu01:pipi} directly.

\section{Analytic calculations in the pQCD approach} \label{sec:2}

\begin{figure}[t,b]
\vspace{-0.5cm} \centerline{\epsfxsize=13 cm \epsffile{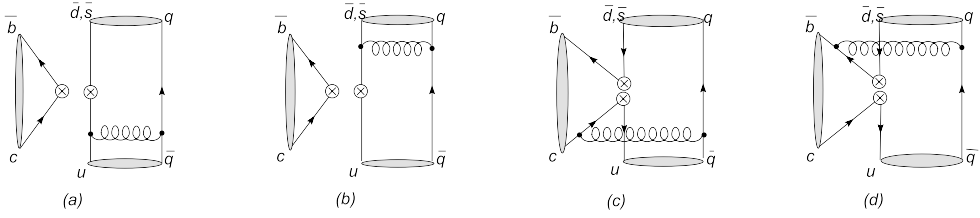}}
\vspace{0.2cm} \caption{Typical Feynman diagrams for the charmless hadronic $B_c \to AP$
decays.}
 \label{fig:fig1}
\end{figure}

In this section, we will calculate the decay amplitudes for 32
charmless hadronic $B_c \to AP/PA$ decays. Analogous to
$B_c \to PV/VP$ decays in Ref.~\cite{Liu09:bcpv},
there are four kinds of annihilation Feynman diagrams contributing to these
considered decays, as illustrated in Fig.~\ref{fig:fig1}.
By analytical evaluation of the two factorizable annihilation ($fa$) diagrams Fig.~1(a) and 1(b),
we find the corresponding decay amplitude
\beq
F_{fa}^{AP} &=&- 8 \pi C_F m_{B_c}^2 \int_0^1 d x_{2} dx_{3}\,\int_{0}^{\infty} b_2 db_2 b_3 db_3\, \non && \times
\left\{h_{fa}(1-x_{3},x_{2},b_{3},b_{2})E_{fa}(t_{a}) \left[x_{2}\phi_{A}(x_2)\phi^A_{P}(x_3)+2 r_A r_0^P \phi_{P}^P(x_3)
\right.\right. \non && \left.\left. \times \left((x_2 + 1)\phi^{s}_{A}(x_2)+ (x_2 -
1)\phi^{t}_{A}(x_{2})\right)\right]+h_{fa}(x_{2},1-x_{3},b_{2},b_{3})E_{fa}(t_{b})\right.\non && \left.\times\left[ (x_3
-1) \phi_{A}(x_2) \phi_{P}^A(x_3) + 2 r_A r_0^P \phi_{A}^s(x_2)\left( (x_3 -2)\phi_{P}^P(x_3)- x_3 \phi_{P}^T(x_3)\right)\right]
\right\}\;, \label{eq:ab}
\eeq
where $\phi_{A}$, $\phi_{A}^{s,t}$ and $\phi_P^{A,P,T}$ denote the distribution amplitudes
of the axial-vector and pseudoscalar mesons,
$r_0^P = m_0^P/m_{B_c}$ with $m_0^P$ standing for the
 chiral scale of pseudoscalar meson($P$), and $C_F=4/3$ is a color factor.
In Eq.~(\ref{eq:ab}), the terms
proportional to $(r_A(r_0^P))^2$ have been neglected because
they are small: less than $7\%$ numerically.
The function $h_{fa}$, the scales $t_i$ and $E_{fa}(t)$ can be found
in Appendix~B of Ref.~\cite{Liu09:bcpv}.

For the two nonfactorizable annihilation ($na$) diagrams Fig.1(c) and 1(d), all three meson wave
functions are involved. The integration of $b_3$ can be performed
using $\delta$ function $\delta(b_3-b_2)$, leaving only integration
of $b_1$ and $b_2$. The corresponding decay amplitude is
\beq
 M_{na}^{AP} &=& -\frac{16 \sqrt{6}}{3}\pi C_F m_{B_c}^2\int_{0}^{1}d x_{2}\,d x_{3}\,\int_{0}^{\infty} b_1d b_1 b_2d b_2\,
 \non && \times \left\{h_{na}^{c}(x_2,x_3,b_1,b_2) E_{na}(t_c)\left[(r_c - x_3 +1) \phi_{A}(x_2)\phi_{P}^A (x_{3})+ r_A
 r_0^P\left(\phi_{A}^s(x_2) \right.\right. \right.\non && \left. \left.\left. \times ((3 r_c + x_2 -x_3 +1) \phi_{P}^P(x_3)
 -(r_c -x_2 -x_3+1)\phi_{P}^T(x_3))+\phi_{A}^t(x_2)\right.\right.\right.\non &&\left.\left. \left. \times ((r_c-x_2 -x_3 +1)
  \phi_{P}^P(x_3)+(r_c-x_2 +x_3 -1)\phi_{P}^T(x_3))\right)\right]-E_{na}(t_d) \right.\non && \left.
 \times\left[ (r_b + r_c +x_2 -1) \phi_{A}(x_2) \phi_{P}^A(x_3) + r_A r_0^P \left(\phi_{A}^s(x_2)((4 r_b
 +r_c +x_2 -x_3\right.\right.\right.\non && \left.\left.\left. -1)\phi_{P}^P(x_3)-(r_c + x_2 +x_3 -1)\phi_{P}^T(x_3))
+\phi_{A}^t(x_2)((r_c + x_2 +x_3 -1)\right.\right.\right. \non && \left.\left.\left. \times\phi_{P}^P(x_3)-(r_c +x_2 -x_3 -1)
\phi_{P}^T(x_3))\right)\right] h_{na}^{d}(x_2,x_3,b_1,b_2)\right\}\;,\label{eq:cd}
\eeq
where $r_{b}= m_{b}/m_{B_c}$, $r_c= m_c/m_{B_c}$ and $r_b +r_c \approx 1$ for $B_c$ meson.

By exchanging the position of the final state mesons $A$ and $P$, we can obtain the
phenomenological topology for $B_c \to P A$ decays easily.
The corresponding decay amplitudes for this type of decay channels can be obtained
directly by the following replacements in Eqs.~(\ref{eq:ab}) and~ (\ref{eq:cd}),
\beq
 \phi_A &\longleftrightarrow& \phi_P^A, \quad  \phi_A^s \longleftrightarrow \phi_P^P,
  \quad \phi_A^t \longleftrightarrow \phi_P^T, \quad  r_A \longleftrightarrow r_0^P .\label{eq:rep}
\eeq

Before we put the things together to write down the decay amplitudes for the
studied decay modes, we give a brief discussion about the $K_{1A}$-$K_{1B}$,
$f_1$-$f_8$ and $h_1$-$h_8$ mixing.

The physical states $K_1(1270)$ and $K_1(1400)$ are the mixtures of the
$K_{1A}$ and $K_{1B}$. $K_{1A}$ and $K_{1B}$ are not mass eigenstates, and
can be mixed together due to the strange and nonstrange light quark mass
difference. The mixing of $K_{1A}$ and $K_{1B}$ can be written as
\beq
|K_1(1270)\rangle&=&|K_{1A}\rangle
{\rm{sin}}\theta_K+|K_{1B}\rangle{\rm{cos}}\theta_K,\\
|K_1(1400)\rangle&=&|K_{1A}\rangle
{\rm{cos}}\theta_K-|K_{1B}\rangle{\rm{sin}}\theta_K.
\eeq
If the SU(3) flavor symmetry between $(u,d,s)$ quark was an exact symmetry,
$K_{1A}$ and $K_{1B}$ would not be mixed  with each other.
As mentioned in the introduction, the mixing angle $\theta_K$ still not be
well determined because of the poor experimental data.
In this paper, for simplicity, we will adopt two reference values as that used in
Ref.~\cite{Yang07:twist}: $\theta_K=\pm 45^\circ$.

Analogous to the $\eta$-$\eta'$ mixing in the pseudoscalar sector,
$f_1(1285)$ and $f_1(1420)$ (the $1^3P_1$ states) will mix in the form of
\beq
\left ( \begin{array}{l} f_1(1285)\\ f_1(1420)\\ \end{array} \right ) =
\left ( \begin{array}{rr}
\cos\theta_3 & \sin\theta_3\\ -\sin\theta_3& \cos\theta_3\\ \end{array} \right )
\left ( \begin{array}{l} f_1\\ f_8\\ \end{array} \right )
\label{eq:f1mixing}
\eeq

Likewise, the $h_1(1170)$ and $h_1(1380)$ ($1^1P_1$ states) system
can be mixed in terms of the pure singlet $|h_1\rangle$ and octet $|h_8\rangle$,
\beq
\left ( \begin{array}{l} h_1(1170)\\ h_1(1380)\\ \end{array} \right ) =
\left ( \begin{array}{rr}
\cos\theta_1 & \sin\theta_1\\ -\sin\theta_1& \cos\theta_1\\ \end{array} \right )
\left ( \begin{array}{l} h_1\\ h_8\\ \end{array} \right )
\label{eq:h1mixing}
\eeq
where the component of $|f_1\rangle, |h_1\rangle$ and $|f_8\rangle, |h_8\rangle$
can be written as
\beq
|f_1\rangle,|h_1\rangle &=&\frac{1}{\sqrt{3}} \left ( |\bar q q \rangle
+ |\bar ss\rangle\right ), \non
|f_8\rangle, |h_8\rangle &=& \frac{1}{\sqrt{6}}\left (|\bar qq\rangle
-2 |\bar ss\rangle \right ),
\eeq
where $q=(u,d)$. The values of the mixing angles for $1^3P_1$ and $1^1P_1$ states
are chosen as \cite{Yang07:twist}:
\beq
\theta_3=38^\circ \quad or \quad 50^\circ; \qquad
\theta_1=10^\circ \quad or \quad 45^\circ.
\eeq

By putting all things together, we can write down the general expression of the
total decay amplitude  for the considered decays:
\beq
{\cal A}(B_c \to  AP) &=& V_{cb}^* V_{uD}
\left\{f_{B_c} F_{fa}^{AP/(PA)}a_1 + M_{na}^{AP/(PA)} C_1 \right\}
\; , \label{eq:amt}
\eeq
where $a_1=C_1/3+C_2$.
Now it is straight forward to present the explicit expressions
of the decay amplitudes for all 32 considered $B_c\to AP$ decays.

\begin{itemize}

\item[]{(1)} For $\Delta S =0$ processes,
\beq
{\cal A}(B_c \to \pi^+  a_1^0)&=&  V_{cb}^* V_{ud} \left\{\left [f_{B_c} F_{fa}^{\pi a_{1u}^0}a_1
+ M_{na}^{\pi a_{1u}^0} C_1 \right ]
  \right. \non && \left.
-\left [f_{B_c}F_{fa}^{a_{1d}^0\pi}a_1 + M_{na}^{a_{1d}^0\pi }
C_1 \right ] \right\}/\sqrt{2}\;,\label{eq:pipa10} \\
{\cal A}(B_c \to a_1^+ \pi^0) &=&  - {\cal A}(B_c \to \pi^+  a_1^0)
= V_{cb}^* V_{ud} \left\{\left [f_{B_c} F_{fa}^{a_1\pi_{u}^0 }
a_1 + M_{na}^{a_1 \pi_{u}^0} C_1\right ]
 \right. \non && \left.
- \left [f_{B_c} F_{fa}^{\pi_{d}^0 a_1 }a_1 + M_{na}^{\pi_{d}^0 a_1 }
C_1\right ]\right\}/\sqrt{2} \;,\label{eq:pi0a1p}\\
{\cal A}(B_c \to a_1^+ \eta) &=& V_{cb}^* V_{ud} \cos \phi
\left\{\left [f_{B_c} F_{fa}^{a_1 \eta_{u} }a_1 + M_{na}^{a_1 \eta_{u}} C_1\right ]
 \right. \non && \left.
+ \left [f_{B_c} F_{fa}^{\eta_{d} a_1 }a_1 + M_{na}^{\eta_{d} a_1 } C_1\right ]
\right\}/\sqrt{2}\;,\\
{\cal A}(B_c \to a_1^+ \eta') &=& V_{cb}^* V_{ud} \sin \phi \left\{\left [f_{B_c}
F_{fa}^{a_1\eta_{u} }a_1 + M_{na}^{a_1 \eta_{u}} C_1\right ]
 \right. \non && \left.
+ \left [f_{B_c} F_{fa}^{\eta_{d} a_1 }a_1 + M_{na}^{\eta_{d} a_1 }
C_1\right ]\right\}/\sqrt{2}\;,
\eeq
\beq
{\cal A}(B_c \to \pi^+  b_1^0)&=&  V_{cb}^* V_{ud} \left\{\left [f_{B_c} F_{fa}^{\pi
b_{1u}^0}a_1 + M_{na}^{\pi b_{1u}^0} C_1 \right ]
  \right. \non && \left.
-\left [f_{B_c}F_{fa}^{b_{1 d}^0\pi}a_1 + M_{na}^{b_{1 d}^0\pi } C_1 \right ]
\right\}/\sqrt{2}\;,\label{eq:pib1} \\
{\cal A}(B_c \to b_1^+ \pi^0) &=&  - {\cal A}(B_c \to \pi^+  b_1^0)=
V_{cb}^* V_{ud} \left\{\left [f_{B_c} F_{fa}^{b_1
\pi_{u}^0 }a_1 + M_{na}^{b_1 \pi_{u}^0} C_1\right ]
 \right. \non && \left.
- \left [f_{B_c} F_{fa}^{\pi_{d}^0 b_1 }a_1 + M_{na}^{\pi_{d}^0 b_1 }
C_1\right ]\right\}/\sqrt{2}\;,\label{eq:pi0b1}\\
{\cal A}(B_c \to b_1^+ \eta) &=& V_{cb}^* V_{ud} \cos \phi \left\{\left [f_{B_c} F_{fa}^{b_1
\eta_{u} }a_1 + M_{na}^{b_1 \eta_{u}} C_1\right ]
 \right. \non && \left.
+ \left [f_{B_c} F_{fa}^{\eta_{d} b_1 }a_1 + M_{na}^{\eta_{d} b_1}
C_1\right ]\right\}/\sqrt{2}\;,\\
{\cal A}(B_c \to b_1^+ \eta') &=& V_{cb}^* V_{ud} \sin \phi \left\{\left [f_{B_c} F_{fa}^{b_1
\eta_{u} }a_1 + M_{na}^{b_1 \eta_{u}} C_1\right ]
 \right. \non && \left.
+ \left [f_{B_c} F_{fa}^{\eta_{d} b_1 }a_1 + M_{na}^{\eta_{d} b_1 }
C_1\right ]\right\}/\sqrt{2}\;,
\eeq
\beq
{\cal A}(B_c \to \pi^+  f_1(1285))&=&  V_{cb}^* V_{ud}
\left\{\frac{\cos\theta_{3}}{\sqrt{3}}
\left [f_{B_c} (F_{fa}^{\pi
f_{1}^{u}}+F_{fa}^{f_{1}^{d}\pi})a_1 \right. \right.\non && \left.\left.
+ (M_{na}^{\pi f_{1}^{u}}+M_{na}^{f_{1}^{d}\pi}) C_1 \right ]
 +\frac{\sin\theta_{3}}{\sqrt{6}}\left [f_{B_c}(F_{fa}^{\pi f_{8}^{u}}
\right. \right.\non && \left.\left.+F_{fa}^{f_{8}^{d}\pi}) a_1 + (M_{na}^{\pi
f_{8}^{u}}+M_{na}^{f_{8}^{d}\pi }) C_1 \right ] \right\}\;,\\
{\cal A}(B_c \to \pi^+  f_1(1420))&=&  V_{cb}^* V_{ud}
\left\{\frac{-\sin\theta_{3}}{ \sqrt{3}}\left [f_{B_c} (F_{fa}^{\pi
f_{1}^{u}}+F_{fa}^{f_{1}^{d}\pi})a_1 \right. \right. \non && \left.\left.
+ (M_{na}^{\pi f_{1}^{u}}+M_{na}^{f_{1}^{d}\pi}) C_1 \right ]
 +\frac{\cos\theta_{3}}{\sqrt{6}}\left [f_{B_c}(F_{fa}^{\pi f_{8}^{u}}
\right. \right.\non && \left. \left.
+F_{fa}^{f_{8}^{d}\pi}) a_1
+ (M_{na}^{\pi f_{8}^{u}}+M_{na}^{f_{8}^{d}\pi }) C_1 \right ] \right\}\;,
\eeq
\beq
{\cal A}(B_c \to \pi^+  h_1(1170))&=&  V_{cb}^* V_{ud}
\left\{\frac{\cos\theta_{1}}{\sqrt{3}}\left [f_{B_c} (F_{fa}^{\pi
h_{1}^{u}}+F_{fa}^{h_{1}^{d}\pi})a_1
\right.\right. \non && \left.\left.
+ (M_{na}^{\pi h_{1}^{u}}+M_{na}^{h_{1}^{d}\pi}) C_1 \right ]
 +\frac{\sin\theta_{1}}{\sqrt{6}}\left [f_{B_c}(F_{fa}^{\pi h_{8}^{u}}
\right.\right. \non && \left. \left.
+F_{fa}^{h_{8}^{d}\pi}) a_1 + (M_{na}^{\pi h_{8}^{u}}+M_{na}^{h_{8}^{d}\pi }) C_1 \right ] \right\}\;,\\
{\cal A}(B_c \to \pi^+  h_1(1380))&=&  V_{cb}^* V_{ud}
\left\{\frac{-\sin\theta_{1}}{\sqrt{3}}\left [f_{B_c} (F_{fa}^{\pi
h_{1}^{u}}+F_{fa}^{h_{1}^{d}\pi})a_1
\right. \right. \non && \left. \left.
+ (M_{na}^{\pi h_{1}^{u}}+M_{na}^{h_{1}^{d}\pi}) C_1 \right ]
 +\frac{\cos\theta_{1}}{\sqrt{6}}\left [f_{B_c}(F_{fa}^{\pi h_{8}^{u}}
\right.\right. \non && \left.\left.
+F_{fa}^{h_{8}^{d}\pi}) a_1 + (M_{na}^{\pi h_{8}^{u}}+M_{na}^{h_{8}^{d}\pi }) C_1 \right ] \right\}\;,
\eeq
\beq
{\cal A}(B_c \to \ov{K}^0 K_1(1270)^+) &=& V_{cb}^* V_{ud}\left\{\sin \theta_{K}
\left [f_{B_c} F_{fa}^{\ov{K}^0 K_{1A}}a_1 +
M_{na}^{\ov{K}^0 K_{1A}} C_1 \right ] \right. \non && \left.
+\cos \theta_{K} \left [f_{B_c} F_{fa}^{\ov{K}^0 K_{1B}}a_1 +
M_{na}^{\ov{K}^0 K_{1B}} C_1 \right ] \right\}\;,\label{eq:kbk127}\\
{\cal A}(B_c \to \ov{K}^0 K_1(1400)^+) &=& V_{cb}^* V_{ud}\left\{\cos \theta_{K}
 \left [f_{B_c} F_{fa}^{\ov{K}^0 K_{1A}}a_1 +
M_{na}^{\ov{K}^0 K_{1A}} C_1 \right ] \right. \non && \left.
-\sin \theta_{K}\left [f_{B_c} F_{fa}^{\ov{K}^0 K_{1B}}a_1 +
M_{na}^{\ov{K}^0 K_{1B}} C_1 \right ] \right\}\;,\label{eq:kbk140}
\eeq

\beq
{\cal A}(B_c \to \ov{K}_1(1270)^0 K^+) &=& V_{cb}^* V_{ud}\left\{\sin \theta_{K}\left [f_{B_c} F_{fa}^{\ov{K}_{1A}^0 K}a_1 +
M_{na}^{\ov{K}_{1A}^0 K} C_1\right ]
\right. \non && \left.
+\cos \theta_{K}\left [f_{B_c} F_{fa}^{\ov{K}_{1B}^0 K}a_1 +
M_{na}^{\ov{K}_{1B}^0 K} C_1 \right ] \right\}\;,\label{eq:k127bkp}\\
{\cal A}(B_c \to \ov{K}_1(1400)^0 K^+) &=& V_{cb}^* V_{ud}\left\{\cos \theta_{K}\left [f_{B_c} F_{fa}^{\ov{K}_{1A}^0 K}a_1 +
M_{na}^{\ov{K}_{1A}^0 K} C_1\right ]
\right. \non && \left.
-\sin \theta_{K}\left [f_{B_c} F_{fa}^{\ov{K}_{1B}^0 K}a_1 +
M_{na}^{\ov{K}_{1B}^0 K} C_1 \right ] \right\}\;. \label{eq:k140bkp}
\eeq


\item[]{(2)} For $\Delta S =1$ processes,
\beq
{\cal A}(B_c \to  K^0 a_1^+) &=&  \sqrt{2}{\cal A}(B_c \to  K^+ a_1^0)
= V_{cb}^* V_{us} \left\{f_{B_c}
F_{fa}^{K^0 a_1} a_1 + M_{na}^{K^0 a_1} C_1 \right\},\; \\
{\cal A}(B_c \to  K^0 b_1^+) &=&  \sqrt{2}{\cal A}(B_c \to  K^+ b_1^0)
= V_{cb}^* V_{us} \left\{f_{B_c}
F_{fa}^{K^0 b_1} a_1 + M_{na}^{K^0 b_1} C_1 \right\},\;
\eeq
\beq
{\cal A}(B_c \to  K_1(1270)^0 \pi^+) &=& \sqrt{2} {\cal A}(B_c \to  K_1(1270)^+ \pi^0)\non
&=& V_{cb}^* V_{us} \left\{\sin \theta_{K}\left [f_{B_c}
F_{fa}^{K_{1A}^{0}\pi}a_1 + M_{na}^{K_{1A}^{0}\pi} C_1\right ] \right. \non && \left. + \cos \theta_{K} \left [f_{B_c}
F_{fa}^{K_{1B}^{0}\pi}a_1 + M_{na}^{K_{1B}^{0}\pi} C_1\right ] \right\},
\label{eq:k127pi}\\
{\cal A}(B_c \to  K_1(1400)^0 \pi^+) &=& \sqrt{2} {\cal A}(B_c \to  K_1(1400)^+ \pi^0)
\non
&=& V_{cb}^* V_{us} \left\{\cos \theta_{K}\left [f_{B_c}
F_{fa}^{K_{1A}^{0}\pi}a_1 + M_{na}^{K_{1A}^{0}\pi} C_1\right ] \right. \non && \left.
- \sin \theta_{K} \left [f_{B_c}
F_{fa}^{K_{1B}^{0}\pi}a_1 + M_{na}^{K_{1B}^{0}\pi} C_1\right ] \right\},
\label{eq:k140pi}
\eeq
\beq
{\cal A}(B_c \to K^+  f_1(1285))&=&
V_{cb}^* V_{us} \left\{ \frac{\cos\theta_{3}}{\sqrt{3}}
\left [f_{B_c} (F_{fa}^{K
f_{1}^{u}}+F_{fa}^{f_{1}^{s} K})a_1
\right. \right.\non && \left.\left.
+ (M_{na}^{K f_{1}^{u}}+M_{na}^{f_{1}^{s} K}) C_1 \right ]
 +\frac{\sin\theta_{3}}{\sqrt{6}}\left [f_{B_c}(F_{fa}^{K f_{8}^{u}}
\right.\right. \non && \left.\left.
-2 F_{fa}^{f_{8}^{s} K}) a_1 + (M_{na}^{K f_{8}^{u}}-2 M_{na}^{f_{8}^{s} K }) C_1 \right ] \right\}\;,\\
{\cal A}(B_c \to K^+  f_1(1420))&=&  V_{cb}^* V_{us} \left\{\frac{-\sin\theta_{3}}{
\sqrt{3}}\left [f_{B_c} (F_{fa}^{K
f_{1}^{u}}+F_{fa}^{f_{1}^{s} K})a_1
\right.\right. \non && \left.\left.
+ (M_{na}^{K f_{1}^{u}}+M_{na}^{f_{1}^{s} K}) C_1 \right ]
 +\frac{\cos\theta_{3}}{\sqrt{6}}\left [f_{B_c}(F_{fa}^{K f_{8}^{u}}
\right.\right. \non && \left.\left.
- 2 F_{fa}^{f_{8}^{s} K}) a_1 + (M_{na}^{K f_{8}^{u}}-2 M_{na}^{f_{8}^{s} K }) C_1 \right ] \right\}\;,
\eeq
\beq
{\cal A}(B_c \to K^+  h_1(1170))&=&  V_{cb}^* V_{us}
\left\{\frac{\cos\theta_{1}}{\sqrt{3}}
\left [f_{B_c} (F_{fa}^{K
h_{1}^{u}}+F_{fa}^{h_{1}^{s} K})a_1
\right.\right. \non && \left.\left.
+ (M_{na}^{K h_{1}^{u}}+M_{na}^{h_{1}^{s} K}) C_1 \right ]
 +\frac{\sin\theta_{1}}{\sqrt{6}}\left [f_{B_c}(F_{fa}^{K h_{8}^{u}}
\right.\right. \non && \left.\left.
-2 F_{fa}^{h_{8}^{s} K}) a_1 + (M_{na}^{K h_{8}^{u}}-2 M_{na}^{h_{8}^{s} K }) C_1 \right ] \right\}\;,\\
{\cal A}(B_c \to K^+  h_1(1380))&=&  V_{cb}^* V_{us}
\left\{\frac{-\sin\theta_{1}}{\sqrt{3}}
\left [f_{B_c} (F_{fa}^{K h_{1}^{u}}+F_{fa}^{h_{1}^{s} K})a_1
\right. \right. \non && \left. \left.
+ (M_{na}^{K h_{1}^{u}}+M_{na}^{h_{1}^{s} K}) C_1 \right ]
 +\frac{\cos\theta_{1}}{\sqrt{6}}\left [f_{B_c}(F_{fa}^{K h_{8}^{u}}
\right.\right. \non && \left.\left.
-2 F_{fa}^{h_{8}^{s} K}) a_1 + (M_{na}^{K h_{8}^{u}}-2 M_{na}^{h_{8}^{s} K }) C_1 \right ] \right\}\;,
\eeq
\beq
{\cal A}(B_c \to K_1(1270)^+  \eta)&=&  V_{cb}^* V_{us} \left\{\sin\theta_{K}
\left [f_{B_c} (\cos\phi F_{fa}^{K_{1A}
\eta_{q}}-\sin\phi F_{fa}^{\eta_{s} K_{1A}})a_1
\right.\right. \non && \left.\left.
+ (\cos\phi M_{na}^{K_{1A} \eta_{q}}- \sin\phi M_{na}^{\eta_{s} K_{1A}}) C_1 \right ]
 \right. \non && \left.
+\cos\theta_{K} \left [f_{B_c}(\cos\phi F_{fa}^{K_{1B} \eta_{q}}
-\sin\phi F_{fa}^{\eta_{s} K_{1B}}) a_1
\right.\right. \non && \left.\left.
+ (\cos\phi M_{na}^{K_{1B} \eta_{q}}-\sin\phi M_{na}^{\eta_{s} K_{1B} }) C_1
\right ] \right\}\;,\label{eq:k127eta}\\
{\cal A}(B_c \to K_1(1400)^+  \eta)&=&  V_{cb}^* V_{us} \left\{\cos\theta_{K}
\left [f_{B_c} (\cos\phi F_{fa}^{K_{1A}
\eta_{q}}-\sin\phi F_{fa}^{\eta_{s} K_{1A}})a_1
\right.\right. \non && \left. \left.
+ (\cos\phi M_{na}^{K_{1A} \eta_{q}}- \sin\phi M_{na}^{\eta_{s} K_{1A}}) C_1 \right ]
 \right. \non &&  \left.
 -\sin\theta_{K} \left [f_{B_c}(\cos\phi F_{fa}^{K_{1B} \eta_{q}}-\sin\phi F_{fa}^{\eta_{s} K_{1B}}) a_1
\right.\right. \non && \left. \left. + (\cos\phi M_{na}^{K_{1B} \eta_{q}}-\sin\phi M_{na}^{\eta_{s} K_{1B} }) C_1 \right ] \right\}\;,\label{eq:k140eta}
\eeq
\beq
{\cal A}(B_c \to K_1(1270)^+  \eta')&=&  V_{cb}^* V_{us} \left\{\sin\theta_{K}
\left [f_{B_c} (\sin\phi F_{fa}^{K_{1A}
\eta_{q}}+ \cos\phi F_{fa}^{\eta_{s} K_{1A}})a_1
\right.\right. \non && \left. \left.
+ (\sin\phi M_{na}^{K_{1A} \eta_{q}}+ \cos\phi M_{na}^{\eta_{s} K_{1A}}) C_1 \right ]
\right. \non &&\left.
+\cos\theta_{K} \left [f_{B_c}(\sin\phi F_{fa}^{K_{1B} \eta_{q}}+ \cos\phi F_{fa}^{\eta_{s} K_{1B}}) a_1
\right.\right. \non && \left. \left.
+ (\sin\phi M_{na}^{K_{1B} \eta_{q}}+ \cos\phi M_{na}^{\eta_{s} K_{1B} }) C_1 \right ] \right\}\;,\label{eq:k127eta'}\\
{\cal A}(B_c \to K_1(1400)^+  \eta')&=&  V_{cb}^* V_{us} \left\{\cos\theta_{K}
\left [f_{B_c} (\sin\phi F_{fa}^{K_{1A}
\eta_{q}}+ \cos\phi F_{fa}^{\eta_{s} K_{1A}})a_1
\right. \right.\non && \left. \left.
+ (\sin\phi M_{na}^{K_{1A} \eta_{q}}+ \cos\phi M_{na}^{\eta_{s} K_{1A}}) C_1 \right ]
\right. \non &&  \left.
-\sin\theta_{K} \left [f_{B_c}(\sin\phi F_{fa}^{K_{1B} \eta_{q}}+ \cos\phi F_{fa}^{\eta_{s} K_{1B}}) a_1
\right.\right. \non && \left. \left. + (\sin\phi M_{na}^{K_{1B} \eta_{q}}+ \cos\phi M_{na}^{\eta_{s} K_{1B} }) C_1 \right ] \right\}\;.
\label{eq:k140eta'}
\eeq

\end{itemize}

\section{Numerical Results and Discussions}\label{sec:3}

In this section, we will calculate the branching ratios for those considered 32
charmless hadronic $B_c \to AP$ decay modes. The input parameters and the wave
functions to be used are given in Appendix \ref{sec:app1}. In
numerical calculations, central values of input parameters will be
used implicitly unless otherwise stated.

For $B_c \to AP$ decays, the decay rate can be written as
\beq
\Gamma =\frac{G_{F}^{2}m^{3}_{B_c}}{32 \pi  } (1- r_A^2) |{\cal A}(B_c
\to AP)|^2\; \label{eq:br}
\eeq
where 
the corresponding decay amplitudes ${\cal A}$ have been
given explicitly in Eqs.~(\ref{eq:pipa10}-\ref{eq:k140eta'}). With the complete decay amplitudes as given in last section, by employing Eq.~(\ref{eq:br}) and
the input parameters and wave functions as given in Appendix~
\ref{sec:app1}, we calculate and present the pQCD predictions for the
CP-averaged branching ratios of the considered decays with errors,
as shown in  Tables~\ref{tab:a1-b1}-\ref{tab:h1-h8}.
The dominant errors come from the uncertainties of  charm quark mass
$m_c=1.5 \pm 0.15$ GeV, the combined Gegenbauer moments $a_i$ of the relevant
meson distribution amplitudes,  and the chiral enhancement factors
$m_0^{\pi}=1.4 \pm 0.3$ GeV  and $m_0^{K}=1.6 \pm 0.1$ GeV, respectively.

\begin{table}[b]
\caption{The pQCD predictions of branching ratios(BRs) for
$B_c \to (a_1, b_1) P$ decays. The source of the dominant errors
is explained in the text.} \label{tab:a1-b1}
\begin{center}\vspace{-0.6cm}
\begin{tabular}[t]{l|l|l|l} \hline  \hline
 $\Delta S =0 $  &                               & $\Delta S =0 $  &   \\
Decay modes      &  BRs$(10^{-7})$              & Decay modes     &   BRs$(10^{-6})$  \\
\hline
 $\rm{B_c \to \pi^+ a_1^0}$ &$3.0^{+0.1}_{-0.3}(m_c)^{+2.3}_{-1.7}(a_i)^{+1.5}_{-1.2}(m_0)$&
 $\rm{B_c \to \pi^+ b_1^0}$ &$4.3^{+1.9}_{-1.4}(m_c)^{+1.8}_{-1.5}(a_i)^{+0.0}_{-0.1}(m_0)$  \\
 $\rm{B_c \to a_1^+ \pi^0}$ &$2.9^{+0.1}_{-0.3}(m_c)^{+2.2}_{-1.7}(a_i)^{+1.4}_{-1.2}(m_0)$&
 $\rm{B_c \to b_1^+ \pi^0}$ &$4.3^{+2.0}_{-1.4}(m_c)^{+2.0}_{-1.5}(a_i)^{+0.1}_{-0.2}(m_0)$  \\
 \hline
 $\rm{B_c \to a_1^+ \eta}$ &$6.8^{+2.4}_{-1.2}(m_c)^{+2.7}_{-2.1}(a_i)^{+0.0}_{-0.0}(m_0)$&
 $\rm{B_c \to b_1^+ \eta}$ &$0.6^{+0.3}_{-0.1}(m_c)^{+0.2}_{-0.1}(a_i)^{+0.0}_{-0.0}(m_0)$  \\
 $\rm{B_c \to a_1^+ \eta'}$ &$4.6^{+1.6}_{-0.8}(m_c)^{+1.7}_{-1.4}(a_i)^{+0.0}_{-0.0}(m_0)$&
 $\rm{B_c \to b_1^+ \eta'}$ &$0.4^{+0.2}_{-0.1}(m_c)^{+0.1}_{-0.1}(a_i)^{+0.0}_{-0.0}(m_0)$  \\
 \hline    \hline
$\Delta S =1 $  &                               & $\Delta S =1 $  &    \\
  Decay modes   &  BRs$(10^{-8})$              & Decay modes     &   BRs$(10^{-7})$  \\
\hline
 $\rm{B_c \to a_1^+ K^0}$ &$3.4^{+1.1}_{-1.2}(m_c)^{+3.2}_{-2.3}(a_i)^{+0.6}_{-0.2}(m_0)$&
 $\rm{B_c \to b_1^+ K^0}$ &$5.4^{+0.9}_{-0.9}(m_c)^{+3.2}_{-2.0}(a_i)^{+0.2}_{-0.0}(m_0)$  \\
 $\rm{B_c \to K^+ a_1^0}$ &$1.7^{+0.6}_{-0.6}(m_c)^{+1.6}_{-1.1}(a_i)^{+0.3}_{-0.1}(m_0)$&
 $\rm{B_c \to K^+ b_1^0}$ &$2.7^{+0.5}_{-0.5}(m_c)^{+1.5}_{-1.1}(a_i)^{+0.1}_{-0.0}(m_0)$  \\
 \hline    \hline
\end{tabular}
\end{center}
\end{table}

\begin{table}[]
\caption{Same as Table~\ref{tab:a1-b1} but for $B_c \to (K_1(1270), K_1(1400)) (\pi, K, \eta, \eta')$ decays.}
\label{tab:k1a-k1b}
\begin{center}\vspace{-0.6cm}
\begin{tabular}[t]{l|l|l}
 \hline \hline
 $\Delta S =0$     &    BRs$(10^{-7})$               &    BRs$(10^{-7})$ \\
 Decay modes       &    $\theta_{K}=45^\circ$     &    $\theta_{K}=-45^\circ$   \\
\hline
 $\rm{B_c \to \ov{K}^0 K_1(1270)^{+}}$ &$8.2^{+1.1}_{-0.5}(m_c)^{+16.3}_{-8.1}(a_i)^{+0.0}_{-0.4}(m_0)$
                                             &$17.4^{+3.2}_{-4.1}(m_c)^{+25.2}_{-16.1}(a_i)^{+0.0}_{-1.5}(m_0)$\\
 $\rm{B_c \to \ov{K}^0 K_1(1400)^{+}}$ &$17.3^{+3.1}_{-4.2}(m_c)^{+24.6}_{-16.1}(a_i)^{+0.0}_{-1.6}(m_0)$
                                             &$8.1^{+1.1}_{-0.5}(m_c)^{+16.1}_{-7.9}(a_i)^{+0.0}_{-0.4}(m_0)$\\
  \hline
 $\rm{B_c \to \ov{K}_1(1270)^0 K^{+}}$ &$15.8^{+7.1}_{-3.3}(m_c)^{+15.6}_{-8.1}(a_i)^{+1.6}_{-0.0}(m_0)$
                                             &$32.0^{+14.4}_{-7.3}(m_c)^{+20.2}_{-19.7}(a_i)^{+0.0}_{-2.0}(m_0)$\\
 $\rm{B_c \to \ov{K}_1(1400)^0 K^{+}}$ &$31.7^{+14.3}_{-7.2}(m_c)^{+20.0}_{-19.5}(a_i)^{+0.0}_{-1.9}(m_0)$
                                             &$15.7^{+7.0}_{-3.4}(m_c)^{+15.2}_{-8.1}(a_i)^{+1.6}_{-0.0}(m_0)$\\
  \hline    \hline
 $\Delta S =1$     &    BRs$(10^{-8})$               &    BRs$(10^{-8})$ \\
 Decay modes       &    $\theta_{K}=45^\circ$     &    $\theta_{K}=-45^\circ$   \\
 \hline
 $\rm{B_c \to K_1(1270)^0 \pi^{+}}$ &$6.8^{+5.1}_{-3.3}(m_c)^{+6.5}_{-4.5}(a_i)^{+0.8}_{-1.3}(m_0)$
                                    &$5.9^{+1.5}_{-0.7}(m_c)^{+3.5}_{-1.9}(a_i)^{+0.6}_{-0.1}(m_0)$\\
 $\rm{B_c \to K_1(1400)^0 \pi^{+}}$ &$5.8^{+1.5}_{-0.6}(m_c)^{+3.6}_{-1.8}(a_i)^{+0.6}_{-0.0}(m_0)$
                                    &$6.8^{+5.0}_{-3.3}(m_c)^{+6.3}_{-4.5}(a_i)^{+0.7}_{-1.3}(m_0)$ \\
\hline
 $\rm{B_c \to K_1(1270)^{+} \pi^0}$ &$3.4^{+2.5}_{-1.6}(m_c)^{+3.3}_{-2.2}(a_i)^{+0.4}_{-0.6}(m_0)$
                                    &$3.0^{+0.7}_{-0.4}(m_c)^{+1.7}_{-1.0}(a_i)^{+0.3}_{-0.1}(m_0)$\\
 $\rm{B_c \to K_1(1400)^{+} \pi^0}$ &$2.9^{+0.7}_{-0.3}(m_c)^{+1.8}_{-0.9}(a_i)^{+0.3}_{-0.0}(m_0)$
                                    &$3.4^{+2.5}_{-1.7}(m_c)^{+3.1}_{-2.2}(a_i)^{+0.4}_{-0.7}(m_0)$\\
  \hline  \hline
 $\Delta S =1$     &    BRs$(10^{-8})$               &    BRs$(10^{-8})$ \\
 Decay modes       &    $\theta_{K}=45^\circ$     &    $\theta_{K}=-45^\circ$   \\
\hline
 $\rm{B_c \to K_1(1270)^+ \eta}$  &$16.8^{+5.0}_{-3.6}(m_c)^{+12.1}_{-10.1}(a_i)^{+0.0}_{-0.0}(m_0)$
                                  &$27.2^{+9.0}_{-8.4}(m_c)^{+14.8}_{-12.9}(a_i)^{+0.0}_{-0.0}(m_0)$\\
 $\rm{B_c \to K_1(1400)^+ \eta}$  &$26.9^{+8.9}_{-8.3}(m_c)^{+14.7}_{-12.8}(a_i)^{+0.0}_{-0.0}(m_0)$
                                  &$16.6^{+5.0}_{-3.5}(m_c)^{+12.2}_{-9.9}(a_i)^{+0.0}_{-0.0}(m_0)$\\
\hline
 $\rm{B_c \to K_1(1270)^+ \eta'}$ &$2.7^{+0.4}_{-0.0}(m_c)^{+4.2}_{-2.7}(a_i)^{+0.0}_{-0.0}(m_0)$
                                  &$11.6^{+1.8}_{-2.1}(m_c)^{+5.0}_{-4.2}(a_i)^{+0.0}_{-0.0}(m_0)$\\
 $\rm{B_c \to K_1(1400)^+ \eta'}$ &$11.5^{+1.7}_{-2.1}(m_c)^{+5.0}_{-4.2}(a_i)^{+0.0}_{-0.0}(m_0)$
                                  &$2.7^{+0.4}_{-0.0}(m_c)^{+4.1}_{-2.7}(a_i)^{+0.0}_{-0.0}(m_0)$\\
\hline \hline
\end{tabular}
\end{center}
\end{table}


\begin{table}[]
\caption{Same as Table~\ref{tab:a1-b1} but for $B_c \to (f_1(1285), f_1(1420)) (\pi, K)$ decays.}
\label{tab:f1-f8}
\begin{center}\vspace{-0.6cm}
\begin{tabular}[t]{l|l|l} \hline  \hline
$\Delta S =0$  &    BRs$(10^{-8})$                 &    BRs$(10^{-8})$   \\
Decay modes    &    $\theta_{3}=38^\circ$     &    $\theta_{3}=50^\circ$   \\
\hline
 $\rm{B_c \to \pi^+ f_1(1285)}$ &$52.4^{+9.3}_{-7.4}(m_c)^{+30.2}_{-23.2}(a_i)^{+21.7}_{-19.0}(m_0)$
                                &$44.8^{+7.0}_{-6.7}(m_c)^{+23.6}_{-19.1}(a_i)^{+19.8}_{-16.8}(m_0)$\\
 $\rm{B_c \to \pi^+ f_1(1420)}$ &$8.5^{+0.5}_{-1.4}(m_c)^{+6.0}_{-5.5}(a_i)^{+0.7}_{-2.6}(m_0)$
                                &$16.0^{+2.8}_{-2.0}(m_c)^{+11.5}_{-9.2}(a_i)^{+0.2}_{-1.5}(m_0)$\\
 \hline    \hline
$\Delta S =1$ &    BRs$(10^{-8})$                 &    BRs$(10^{-8})$\\
Decay modes   &    $\theta_{3}=38^\circ$     &    $\theta_{3}=50^\circ$   \\
\hline
 $\rm{B_c \to K^+ f_1(1285)}$ &$1.6^{+1.0}_{-0.7}(m_c)^{+3.4}_{-1.8}(a_i)^{+0.5}_{-0.4}(m_0)$
                              &$1.5^{+1.0}_{-0.5}(m_c)^{+3.9}_{-1.4}(a_i)^{+0.5}_{-0.3}(m_0)$\\
 $\rm{B_c \to K^+ f_1(1420)}$ &$7.4^{+0.3}_{-0.0}(m_c)^{+3.2}_{-2.8}(a_i)^{+0.4}_{-0.2}(m_0)$
                              &$7.5^{+0.3}_{-0.0}(m_c)^{+3.2}_{-3.1}(a_i)^{+0.4}_{-0.4}(m_0)$\\
 \hline \hline
 \end{tabular}
\end{center}
\end{table}

\begin{table}[]
\caption{Same as Table~\ref{tab:a1-b1} but for $B_c \to (h_1(1170), h_1(1380)) (\pi, K)$ decays.}
\label{tab:h1-h8}
\begin{center}\vspace{-0.6cm}
\begin{tabular}[t]{l|l|l}
 \hline \hline
$\Delta S = 0 $        &    BRs$(10^{-8})$                 &    BRs$(10^{-8})$\\
Decay modes            &    $\theta_{1}=10^\circ$     &    $\theta_{1}=45^\circ$   \\
\hline
 $\rm{B_c \to \pi^+ h_1(1170)}$ &$60.5^{+28.6}_{-0.0}(m_c)^{+32.1}_{-16.9}(a_i)^{+25.8}_{-8.0}(m_0)$
                                &$49.1^{+28.8}_{-4.9}(m_c)^{+33.7}_{-10.9}(a_i)^{+20.7}_{-5.5}(m_0)$ \\
 $\rm{B_c \to \pi^+ h_1(1380)}$ &$1.2^{+2.4}_{-1.0}(m_c)^{+3.7}_{-0.2}(a_i)^{+0.5}_{-0.0}(m_0)$
                                &$12.4^{+3.9}_{-0.0}(m_c)^{+3.3}_{-3.7}(a_i)^{+5.5}_{-2.2}(m_0)$  \\
 \hline    \hline
$\Delta S = 1 $        &    BRs$(10^{-8})$                 &    BRs$(10^{-8})$\\
Decay modes            &    $\theta_{1}=10^\circ$     &    $\theta_{1}=45^\circ$   \\
\hline
 $\rm{B_c \to K^+ h_1(1170)}$ &$14.9^{+2.0}_{-1.8}(m_c)^{+12.6}_{-8.0}(a_i)^{+0.0}_{-0.3}(m_0)$
                              &$16.8^{+5.7}_{-4.5}(m_c)^{+6.9}_{-5.7}(a_i)^{+0.3}_{-1.6}(m_0)$  \\
 $\rm{B_c \to K^+ h_1(1380)}$ &$22.2^{+14.5}_{-7.3}(m_c)^{+15.6}_{-13.1}(a_i)^{+0.0}_{-0.2}(m_0)$
                              &$20.2^{+10.6}_{-4.5}(m_c)^{+11.7}_{-8.2}(a_i)^{+0.0}_{-0.8}(m_0)$  \\
\hline \hline
\end{tabular}
\end{center}
\end{table}


Based on the numerical results as given in Tables I -IV, we have the following
remarks:
\begin{itemize}
\item
The pQCD predictions for the CP-averaged branching ratios of considered
$B_c$ decays vary in the range of $10^{-6}$ to  $10^{-8}$.
There is no CP violation for all these decays within the standard model, since there
is only one kind of tree operator involved in the decay amplitude of
all considered $B_c$ decays, which can be seen from Eq.~(\ref{eq:amt}).

\item
Among  the considered $B_c\to AP$ decays, the
pQCD predictions for the branching ratios of those
$\Delta S = 0$ processes are generally much larger than those of $\Delta S =1$
channels (one of the two final state mesons is a strange meson),
the main reason is the enhancement of  the large CKM factor $|V_{ud}/V_{us}|^2 \sim 19$
for those $\Delta S = 0$ decays as generally expected.
For $B_c\to (a_1^+, b_1^+)(\pi^0, K^0)$ decays, however,
the difference is not so large, because the
enhancement due to the CKM factor is partially canceled
by the differences between the magnitude of individual decay amplitude
$|F_{fa}^{a(b)_1^+\pi^0}|$ and $|F_{fa}^{a(b)_1^+ K^0}|$.

\item
For $B_c \to (a_1, b_1) \pi$ decays, the same component of $\bar u u - \bar d d$
is involved in both axial-vector $(a_1^0, b_1^0)$  and the 
pseudoscalar $\pi^0$ meson at the quark level.
We therefore find the same branching ratios for $B_c \to \pi^+ a_1^0$ and
$B_c \to a_1^+ \pi^0$, and for $B_c \to \pi^+ b_1^0$ and $B_c \to b_1^+ \pi^0$,
respectively.

\item
From the numerical results as shown in Table~\ref{tab:a1-b1}, one can see
that:
\beq
Br(B_c \to b_1 \pi) &\sim & 14 \times Br(B_c \to a_1 \pi), \non
Br(B_c \to b_1 K) &\sim & 16 \times Br(B_c \to a_1 K).
\eeq
This pattern agrees well with that as given in  Ref.~\cite{Cheng07:ap,Wang08:a1}.

\item
Unlike $B_c \to (a_1, b_1) (\pi, K)$ decays,  we find that
\beq
Br(B_c \to a_1 (\eta, \eta')) \sim Br(B_c \to b_1 (\eta, \eta')).
\eeq
The main reason is that the suppressed factorizable annihilation amplitudes
cancel each other for $B_c \to a_1 \etap$ decays, while the
enhanced nonfactorizable ones  cancel each other for $B_c \to b_1 \etap$ decays.

\item
For $B_c \to \ov K^0 (K_1(1270)^+, K_1(1400)^+)$ and $B_c\to
(\ov K_1(1270)^0, \ov K_1(1400)^0) K^+$ decays, their BRs strongly
depend on the value of the mixing angle $\theta_K$ of the $K_{1A}$-$K_{1B}$ system.
From Table II, one can see that
\beq
\frac{Br(B_c \to \ov K^0 K_1(1400)^+)}{Br(B_c \to \ov K^0 K_1(1270)^+)}
\approx
\frac{Br(B_c \to  K^+ \ov K_1(1400)^0)}{Br(B_c \to K^+ \ov K_1(1270)^0)} \approx 2,\label{eq:ratio1}
\eeq
for $\theta_K=45^\circ$, while
\beq
\frac{Br(B_c \to \ov K^0 K_1(1400)^+)}{Br(B_c \to \ov K^0 K_1(1270)^+)}
\approx
\frac{Br(B_c \to  K^+ \ov K_1(1400)^0)}{Br(B_c \to K^+ \ov K_1(1270)^0)} \approx \frac{1}{2},\label{eq:ratio2}
\eeq
for $\theta_K=- 45^\circ$. This means that one can determine the sign and  size of
$\theta_K$ after enough $B_c$ events become available at the LHC experiment.

\item
For the $\Delta S = 1$ $B_c \to K_1 \pi$ decays, their decay rates have a very
weak  dependence on the value of mixing angle $\theta_K$:
\beq
Br(B_c \to K_1(1270)^0 \pi^+) \approx Br(B_c \to K_1(1400)^0 \pi^+) \approx
6\times 10^{-8}, \\
Br(B_c \to K_1(1270)^+ \pi^0) \approx Br(B_c \to K_1(1400)^+ \pi^0) \approx
3\times 10^{-8},
\eeq
for both $\theta_K=45^\circ$ and $-45^\circ$. This point will also be tested at LHC.

\item
For $B_c \to K_1 \etap $ decays, the pQCD predictions have a strong $\theta_K$
dependence:
\beq
\frac{Br(B_c \to K_1(1400)^+ \eta)}{Br(B_c \to K_1(1270)^+ \eta)} \approx 1.6, \non
\frac{Br(B_c \to K_1(1400)^+ \eta^\prime)}{Br(B_c \to K_1(1270)^+ \eta^\prime)} \approx 4.3,\label{eq:ratio3}
\eeq
for $\theta_K=45^\circ$, while
\beq
\frac{Br(B_c \to K_1(1400)^+ \eta)}{Br(B_c \to K_1(1270)^+ \eta)} \approx \frac{1}{1.6}, \non
\frac{Br(B_c \to K_1(1400)^+ \eta^\prime)}{Br(B_c \to K_1(1270)^+ \eta^\prime)} \approx \frac{1}{4.3},\label{eq:ratio4}
\eeq
for $\theta_K=-45^\circ$. It is easy to see that these $B_c \to K_1 \etap $ decays are
sensitive to the mixing angle $\theta_K$.
Analogous to the $B_c \to K^* (\eta, \eta')$ decays~\cite{Liu09:bcpv}, the above four
decays are dominated by the factorizable annihilation diagrams.

\item
The theoretical predictions for the branching ratios of $B_c \to K_1(1270)P$ and $B_c \to
K_1(1400)P$ for $\theta_K=45^\circ$, as listed in the column two of Table II, are roughly
exchanged with respect to those of the third column for the choice of
$\theta_K=-45^\circ$. Such simple relation comes from the fact that the two states
$K_1(1270)$ and $K_1(1400)$ can go one into another as a mixture of $K_{1A}$ and $K_{1B}$
states when one sets the mixing angle $\theta_K=45^\circ$ or $-45^\circ$ respectively,
as can be seen from Eqs.~(11,12),
\beq
|K_1(1270)\rangle|_{\theta_K=45^\circ}&=& |K_1(1400)\rangle|_{\theta_K=-45^\circ},\non
|K_1(1400)\rangle|_{\theta_K=45^\circ} &=& -|K_1(1270)\rangle|_{\theta_K=-45^\circ}.
\eeq
This relation further leads to the following relations between the decay amplitudes of
$B_c \to K_1 P$:
\beq
{\cal A} (B_c \to K_1(1270) P)|_{\theta_K=45^\circ} &=& {\cal A} (B_c \to K_1(1400) P)|_{\theta_K=-45^\circ}\;,\non
{\cal A} (B_c \to K_1(1400) P)|_{\theta_K=45^\circ} &=&
-{\cal A} (B_c \to K_1(1270) P)|_{\theta_K=-45^\circ}\;,
\label{eq:scheme2}
\eeq
and finally we obtain the special pattern of branching ratios as listed in Table II.
The small differences in corresponding decay rates are due to the difference
in the masses of $K_1(1270)$ and $K_1(1400)$ mesons.
The numerical relations as shown in Eqs.~(\ref{eq:ratio1}-\ref{eq:ratio2})
and~(\ref{eq:ratio3}-\ref{eq:ratio4}) are also induced by the same mechanism.

\item
For the four $B_c \to f_1 (K, \pi)$ decays, one can see from
Table~\ref{tab:f1-f8} that
\beq
\frac{Br(B_c \to \pi^+ f_1(1285))}{Br(B_c \to \pi^+ f_1(1420))} &\approx &
\left\{ \begin{array}{ll}
6.2& {\rm for \ \ \theta_3=38^\circ } \\
2.8& {\rm for \ \ \theta_3=50^\circ} \\ \end{array} \right.
\label{eq:f1a}
\eeq
and
\beq
\frac{Br(B_c \to K^+ f_1(1285))}{Br(B_c \to K^+ f_1(1420))} &\approx & 0.2
\label{eq:f1b}
\eeq
for $\theta_3=38^\circ$ and $50^\circ$.
The relations in Eqs.~(\ref{eq:f1a},\ref{eq:f1b}) can be understood as follows:
(a) since $f_1(1285)$ and $f_1(1420)$ are the mixed states of $f_1$ and $f_8$
(see Eq.~(\ref{eq:f1mixing}) ) and both $\sin\theta_3 $ and $\cos\theta_3$ are positive for
$\theta_3=38^\circ$ and $50^\circ$,
the contribution from the common component $(\bar qq)$
of $f_1$ and $f_8$ will interfere constructively (destructively) for
$B_c \to \pi^+ f_1(1285)$ ($B_c\to  \pi^+ f_1(1420)$) decay, this results in the large
difference for the decay rate of the two decays;
(b) for the two $\Delta S=1$ decays, however, the new component $(\bar ss)$ will
provide additional contributions  to the considered decays.
Furthermore, the contributions from $(\bar ss)$ and $\bar qq$ interfere
constructively for $B_c \to K^+ f_1(1420)$, but destructively for $B_c \to K^+ f_1(1285)$
decay.

\item
The pQCD predictions for $B_c\to (h_1(1170), h_1(1380) (K, \pi)$ decays, as given in
Table IV, can be explained in a similar way as for $B_c\to (f_1(1285), f_1(1400) (K, \pi)$
decays.

\item
Since the LHC experiment can measure the $B_c$ decays with a
branching ratio at $10^{-6}$ level~\cite{ekou09:ncbc}, our pQCD predictions for
the branching ratios of $B_c \to K (K_1(1270), K_1(1400))$ and $b_1 \pi$
decays could be tested in the forthcoming LHC experiments.

\end{itemize}

It is worth stressing that the theoretical predictions in
the pQCD approach still have large theoretical errors induced
by the still large uncertainties of many input parameters, e.g.
Gegenbauer moments $a_i$.
For most considered pure annihilation $B_c$ decays, it is hard to observe them
even in LHC due to their tiny decay rate.
Their observation at LHC, however, would mean  a large non-perturbative
contribution or a signal for new physics beyond the SM.

We here calculated the branching ratios of the pure annihilation $B_c \to AP$ decays
by employing the pQCD approach.
We do not consider the possible long-distance (LD) contributions, such as the
re-scattering effects, although they may be large and affect the theoretical
predictions. It is beyond the  scope of this work.


\section{Summary}\label{sec:sum}

In short, we studied the charmless hadronic $B_c \to AP$
decays by employing the pQCD factorization approach based on the
$k_T$ factorization theorem. These considered decay channels can
occur only via the annihilation diagram in the SM and they will provide an important platform
for testing the magnitude of the annihilation contribution and understanding the content of the
axial-vector mesons.

The pQCD predictions for CP-averaged branching ratios are
displayed in Tables~(\ref{tab:a1-b1}-\ref{tab:h1-h8}).
From our numerical evaluations and phenomenological analysis, we
found the following results:
\begin{itemize}
\item
The pQCD predictions for the branching ratios vary in the
range of $10^{-6}$ to $10^{-8}$. The $B_c\to \ov K^0 (K_1(1270)^+,
K_1(1400)^+)$ and other decays with a decay rate at $10^{-6}$ or
larger  could be measured at the LHC experiment.

\item
For $B_c \to A\;P$ decays, the branching ratios of  $\Delta
S= 0$ processes are generally much larger than those of $\Delta S =1$
ones. Such differences are mainly induced by the CKM factors
involved: $V_{ud}\sim 1 $ for the former decays while $V_{us}\sim
0.22$ for the latter ones.

\item
Since the behavior for $^1P_1$ meson is much different from
that for $^3P_1$ meson, the branching ratios of pure annihilation
$B_c \to A(^1P_1) P$ are basically larger than that of $B_c \to
A(^3P_1) P$, which can be tested in the LHC and Super-B
experiments.

\item
The pQCD predictions about the branching ratios of $B_c \to K_1 \etap $ and $K_1 K$
decays are rather sensitive to the value of the mixing angle $\theta_K$.
One can determine $\theta_K$ through the measurement of these decays if
enough $B_c$ events become available at the LHC experiment.

\item
The pQCD predictions still have large theoretical
uncertainties, mainly induced by the uncertainties of the Gegenbauer
moments $a_i$ in the meson distribution amplitudes.

\item
Because only tree operators are involved, the CP-violating asymmetries for
these considered $B_c$ decays are absent naturally.

\end{itemize}

\begin{acknowledgments}

X.Liu would like to thank Hai-Yang Cheng, Wei Wang,
You-Chang Yang and Run-Hui Li for valuable discussions. This work
is supported by the National Natural Science Foundation of China
under Grant No.10975074 and 10735080, by the
Project on Graduate Students' Education and Innovation of Jiangsu
Province under Grant No. ${\rm CX09B_{-}297Z}$, and by the Project on
Excellent Ph.D Thesis of Nanjing Normal University under Grant
No. 181200000251.

\end{acknowledgments}


\begin{appendix}

\section{Input parameters and distribution amplitudes} \label{sec:app1}

The masses~({\rm GeV}), decay constants~({\rm GeV}), QCD
scale~({\rm GeV}) and $B_c$ meson lifetime~({\rm ps}) are
\beq
 \Lambda_{\overline{\mathrm{MS}}}^{(f=4)} &=& 0.250, \quad m_W = 80.41, \quad  m_{B_c} = 6.286, \quad  f_{B_c} = 0.489,  \non
 m_{a_1} &=& 1.23,\;\;\;\;\quad f_{a_1} = 0.238,\;\quad m_{b_1}=1.21, \;\;\;\;\;\; f_{b_1} = 0.180,
   \non
m_{K_{1A}} &=& 1.32, \quad f_{K_{1A}} = 0.250 \;\; \quad
m_{K_{1B}} =1.34, \;\;\quad f_{K_{1B}}= 0.190,
 \non
   f_{f_1}&=& 0.245, \;\quad m_{f_1}=1.28, \;\;\;\quad f_{f_8}= 0.239,
  \; \quad m_{f_8} =1.28, \non
     f_{h_1}&=& 0.180, \;\quad m_{h_1}=1.23, \;\;\;\quad f_{h_8}= 0.190,
  \; \quad m_{h_8} =1.37, \non
     m^\pi_{0}&=& 1.4,\;\;\;\; \quad m_0^K=1.6,\;\;\;\; \quad m_0^{\eta_q}= 1.08,
  \;\; \quad m_0^{\eta_s} =1.92, \non
    m_b &=& 4.8, \;\;\;\;\;\;\quad f_{\pi}= 0.131, \;\;\quad f_K = 0.16, \;\;\quad
    \tau_{B_c^+}= 0.46\; \;.
 \label{para}
\eeq

For the CKM matrix elements, here we adopt the Wolfenstein
parametrization for the CKM matrix, and take $A=0.814$ and
 $\lambda=0.2257$, $\bar{\rho}=0.135$ and $\bar{\eta}=0.349$ \cite{Amsler08:pdg}.

For the distribution amplitudes of pseudoscalar mesons,
we adopt the same forms as that used in the literature
(See Ref.~\cite{Liu09:bcpv} and references therein).

The twist-2 distribution amplitudes for the longitudinally
polarized axial-vector $^3P_1$ and $^1P_1$ mesons can be
parameterized as~\cite{Yang07:twist,Li09:afm}: \beq
 \phi_A(x) & = & \frac{f}{2 \sqrt{2 N_c}} \left\{ 6 x (1- x) \left[ a_0^\parallel + 3
a_1^\parallel\, t +
a_2^\parallel\, \frac{3}{2} ( 5t^2  - 1 ) \right] \right\},\label{eq:lda}\eeq

As for twist-3 LCDAs, we use the following form:
\begin{eqnarray}
\phi_{A}^t(x) &= &\frac{3 f}{2\sqrt{2N_c}}\left\{ a_0^\perp t^2+ \frac{1}{2}\,a_1^\perp\,t (3 t^2-1) \right\}
 ,\\
\phi_{A}^s(x)&=& \frac{3 f}{2\sqrt{2N_c}}
\frac{d}{dx}\left\{ x (1- x) ( a_0^\perp + a_1^\perp t ) \right\}.
 \end{eqnarray}
 where $f$ is the decay constant and $t = 2 x-1$. It should be noted that for the distribution amplitudes of strange axial-vector
 mesons $K_{1A}$ and $K_{1B}$, $x$ stands for the momentum fraction carrying by $s$ quark.

 Here, the definition of these distribution amplitudes $\phi_A(x)$ satisfy the
 following relation:
 \beq
\int_0^1 \phi_{^3P_1}(x) &=& \frac{f}{2 \sqrt{2 N_c}}, \;\;\;\;\;\;\;\;  \int_0^1 \phi_{^1P_1}(x) = a_0^{||, ^1P_1} \frac{f}{2 \sqrt{2 N_c}}.
 \eeq
 where we have used $a_0^{||, ^3P_1}= 1$.

The Gegenbauer moments have been studied extensively in the
literatures (see Ref.~\cite{Yang07:twist} and references therein), here we adopt the following
values:
\beq
 a_2^{||, a_1}&=& -0.02\pm 0.02;\;\;\;\;\;\;\;\;\;  a_1^{\perp, a_1}= -1.04\pm 0.34;\;\;\;\;\;\;\;\;\;  a_1^{||, b_1}=  -1.95\pm 0.35; \non
 a_2^{||,f_1} &=& -0.04\pm 0.03;\;\;\;\;\;\;\;\;\; a_1^{\perp,f_1}= -1.06\pm 0.36;\;\;\;\;\;\;\;\;\; a_1^{||,h_1}= -2.00\pm 0.35;\non
 a_2^{||,f_8} &=& -0.07\pm 0.04;\;\;\;\;\;\;\;\;\; a_1^{\perp,f_8}= -1.11\pm 0.31;\;\;\;\;\;\;\;\;\; a_1^{||,h_8}= -1.95\pm 0.35;\non
 a_1^{||, K_{1A}}&=& 0.00\pm 0.26;\;\;\;\;\;\;\;\;\; a_2^{||, K_{1A}}= -0.05\pm 0.03;\;\;\;\;\;\;\;\; a_0^{\perp, K_{1A}}= 0.08\pm 0.09;\non
 a_1^{\perp, K_{1A}}&=& -1.08\pm 0.48;\;\;\;\;\;\;\; a_0^{||, K_{1B}}= 0.14\pm 0.15;\;\;\;\;\;\;\;\;\; a_1^{||, K_{1B}}= -1.95\pm 0.45;\non
 a_2^{||, K_{1B}}&=& 0.02\pm 0.10;\;\;\;\;\;\;\;\;\; a_1^{\perp, K_{1B}}=0.17\pm 0.22.
\eeq

\end{appendix}


\end{document}